\documentclass[twocolumn]{aastex63}
\usepackage{commath}
\usepackage{epsfig}

\newcommand{\Ms}{{\ensuremath{\rm M_{\odot} }}}

\shorttitle{magnetic fields}
\shortauthors{Latif  et al.}

\begin{document}

\title{Role of magnetic fields in the formation of direct collapse black holes }



\correspondingauthor{Muhammad A. Latif}
\email{latifne@gmail.com}

\author{Muhammad A. Latif}
\affiliation{Physics Department, College of Science, United Arab Emirates University, PO Box 15551, Al-Ain, UAE}

\author{Dominik R. G. Schleicher}
\affiliation{Astronomy Department, Universidad de Concepci\'on, Barrio Universitario, Concepci\'on, Chile}

\author{Sadegh Khochfar}
\affiliation{Institute for Astronomy, University of Edinburgh, Royal Observatory, Blackford Hill, Edinburgh EH9 3HJ, UK}

%
%
\begin{abstract}
Direct collapse black holes (DCBHs) are the leading candidates for the origin of the first supermassive black holes.  However, the role of magnetic fields during their formation is still unclear as none of the previous studies has been evolved long enough to assess their impact during the accretion phase.  Here, we report the results from a suite of 3D cosmological magneto-hydrodynamic (MHD) simulations which are evolved for 1.6 Myrs comparable to the expected lifetime of supermassive stars (SMSs).  Our findings suggest that magnetic fields are rapidly amplified  by strong accretion shocks irrespective of  the initial magnetic field strength and reach the saturation state. They  stabilize  the accretion disks and significantly reduce fragmentation by enhancing the Jeans mass in comparison with pure hydrodynamical runs. Although the initial clump masses are larger in MHD runs, the rapid  coalescence of clumps in non-MHD cases  due to the higher degree of fragmentation results in similar masses.  Overall, the central clumps have masses of $\rm 10^5~M_{\odot}$  and the mean mass accretion rates of $\rm \sim 0.1 ~M_{\odot}/yr$ are similar in both MHD and non-MHD cases. The multiplicity of SMSs is significantly reduced in MHD simulations.  Such strongly amplified magnetic fields are expected to launch Jets and outflows which may be detected with upcoming radio telescopes.

\end{abstract}


\section{Introduction} \label{sec:intro}
High redshift surveys have detected more than 400 quasars at z$>$5.7 \citep{Fan03,Mort11,Wu15, Ban17,Wang21} powered by supermassive black holes (SMBHs) of about $\rm \sim 10^9 ~M_{\odot}$.  The existence of such objects within the first billion years of the Universe is still a mystery and poses the biggest challenge to our understanding of early structure formation. Various pathways  have  been proposed for the formation of black holes, which from the astrophysical point of a view can be categorized into three mechanisms: 1) remnants of   pop III stars so-called stellar mass black holes, 2) stellar dynamical processes in dense stellar clusters leading to the formation of a massive black hole, 3)  monolithic collapse of a massive protogalactic gas cloud into an intermediate mass black hole commonly known as the direct collapse black holes (DCBH) scenario. Details of these mechanisms can be found in dedicated reviews \citep{Latif16P,Woods19,In19}. Black holes  forming from the core collapse of massive pop III stars at $z \sim 30$ are expected to have about  $\rm \sim 100 ~M_{\odot}$ \citep{Sugi20,Latif22}  and are difficult to grow due to the shallow dark matter potentials of their host  halos \citep{Johnson2007,Alvarez09,Riaz18,Latif20b}. They may require multiple episodes of  super-Eddington accretion to reach the observed masses within the first billion year \citep{Lupi14, Pacucci15,Mayer15, Mayer19}. Runaway stellar collisions  during the core collapse of dense stellar clusters  may result in a very massive star which subsequently collapses into a black hole of  a  few times $\rm 10^3-10^4$ \citep{Devecchi12,Katz2015,Latif16d,Yaji16, sak17,Reinoso18a,Chon20,Das21,Ver21}, see \cite{Sch22} for more details.

The direct collapse scenario yields black holes of $\rm 10^5~M_{\odot}$ which are about two orders of magnitude more massive than in the other scenarios \citep{Latif13c,Sakurai16}.  The key requirement for the DCBH mechanism is a mass inflow rate of $\rm \sim 0.1~M_{\odot}/yr$ \citep{Hosokawa13,Schleicher13}. Such high accretion rates can be more easily obtained  thermodynamically in massive primordial  halos by quenching  $\rm H_2$ formation which results in an isothermal collapse with $\rm T \sim 8000$ K  and larger sound speeds ($\rm \dot{M} \sim {c_s^3}/{G} \sim 0.1~M_{\odot}/yr \left( {T}/{8000~K}\right)^{3/2}$). Alternatively, inflow  rates of $\rm \sim 0.1~M_{\odot}/yr$ can be obtained dynamically via bars within bars instabilities \citep{Begel06} or through the rapid growth of halos via major mergers in combination with moderate UV flux in metal free halos \citep{LatifV15,Wise19}, or through large baryonic streaming motions \citep{Hir17}. Recently, \cite{Latif22N} found that dense cold turbulent flows give birth  to DCBHs without the need of  an external UV flux. Numerical simulations investigating the feasibility of the DCBH mechanism find that large accretion rates of $\rm 0.1-1~M_{\odot}/yr$ can be obtained \citep{Latif13d,Regan19,Latif20}. However, their number density is still an open question due to the dependence of  the required critical UV flux  (for suppressing $\rm H_2$ formation) on local conditions and on the environment  \citep{Sugimura14,Latif15a,Agarwal2015B}.

Previous works mainly focused on hydrodynamical aspects of the DCBH scenario and the role of magnetic fields is not yet fully clear. They are expected to influence DCBH formation by transferring angular momentum, exerting magnetic torques and may even suppress fragmentation by enhancing the Jeans mass. Magnetic fields observed in the local universe are thought to be responsible for various astrophysical phenomena including the launch of magnetic outflows and jets \citep{Beck99,Beck07,Pud12}.  Furthermore, observation of rotation measures in high z quasars up to $z \sim 5.3$ reveal the presence of magnetic fields at earlier times \citep{Murphy09,Hammond12}.  Future radio observatories such as ngVLA and SKA will probe magnetic fields at high redshifts \citep{Arsh09,Nyland18} which are expected to efficiently form via mechanisms such as the small scale dynamo \citep{Sch13,Schb13}.

The standard model does not yield any  constraints on the strength of the initial magnetic fields. They can be generated through  electro-weak or quantum chromodynamics phase transitions during the cosmic inflation or alternatively via  mechanisms such as  the Biermann battery effect and  the Weibel instability (see review by \cite{Wid12}). Irrespective of their origin, the seed magnetic fields are many orders of magnitude smaller than the present day fields. They can be initially strengthened via gravitational compression in the domain of flux freezing and later can get further amplified via the small scale dynamo as it converts turbulent into magnetic energy on short timescales. In the context of  star formation various studies confirm that the small scale dynamo becomes operational and amplifies seed fields on very short times scales \citep{Schleicher10,Sur10,Fed11,Schobera,Turk12,Grete19, Sharda20}. These studies further suggest that such strong fields may impact the IMF of primordial stars. On the other hand, the role of magnetic fields in the formation of DCBHs is not fully clear. \cite{Latif13m} (hereafter L13) and \cite{Grete19} (hereafer G19) performed cosmological magneto-hydrodynamical simulations and have shown  that magnetic fields get amplified very efficiently in atomic cooling halos where gas mainly cools by atomic line radiation.  Furthermore, \cite{LatifMag14} (hereafter L14) found that irrespective of the initial seed field strength, the magnetic field is amplified within the  free-fall time scale  in the presence of strong accretion shocks. However, these simulations could not be evolved long enough to determine the role of such magnetic fields during the late accretion phase of supermassive stars due to the exorbitant computational costs. 

Recently, \cite{Hir21} performed idealized simulations  to study the collapse of atomically cooling magnetized gas cloud.  They found that magnetic fields  enhance the mass accretion  onto the gas cloud which results in a higher degree of fragmentation but simultaneously the coalescence rate of clumps gets increased. Consequently, the massive star grows at a faster rate and thus  magnetic fields support the DCBH formation. However, their simulations  were performed for an isolated gas cloud  assuming a Bonner-Ebbort sphere  and were evolved only for 700 yrs. Therefore, it is unclear how magnetic fields regulate the formation of SMSs over longer times in realistic cosmological environments, what is their impact on degree of fragmentation  and how they  influence the final masses of the resulting objects. To investigate  these questions, we perform 3D MHD  cosmological simulations for three distinct halos and evolve them  for about 1.6 Myrs and compare our results with hydrodynamical simulations. These simulations enable us to investigate the role of the magnetic fields during the accretion phase.

Our article is organized as follows. In section 2, we discuss the numerical methods and the simulation setup. We present our main findings in section 3 and confer our conclusions in section 4.

\section{Methods} \label{sec:methods}

Numerical simulations  are conducted with  the cosmological magnetohydrodynamics open source code  \textsc{ENZO} \citep{Enzo14} which utilizes an adaptive mesh refinement approach to cover a large  range of spatial scales: from Mpc  to AU scales.  Enzo employs the Harten-Lax-van Leer (HLL) Riemann solver with a piece-wise linear construction for magnetohydrodynamics,  the $N$-body particle-mesh technique to compute DM dynamics and a multigrid Poisson solver for self-gravity calculations.  It uses the Dedner scheme \citep{Wang08,Wang10} for divergence cleaning.

Our simulations start  at $z$=150 with cosmological initial  conditions generated from  Gaussian random density perturbations with MUSIC  \citep{Hahn11}  and use the following cosmological parameters from the \textit{Planck} mission: $\Omega_{\mathrm{M}}=$ 0.3089, $\Omega_{\Lambda}=$ 0.691, $\Omega_{\mathrm{b}} = $ 0.0486, $h =$ 0.677, $\sigma_8 = $ 0.816, and $n =$ 0.967 \citep{Planck16}.  We take a comoving periodic box of 1 Mpc/h in size with top grid resolution of $\rm 256^3$ and the same number of DM particles. The most massive halo of a few times $\rm 10^7~M_{\odot}$ in the computational volume is selected at the earliest redshift of its formation. The selected halo is placed at the center of the box and  two additional static grids are added in the central 250 ckpc region yielding an effective DM resolution of $\rm \sim 67~ M_{\odot}$.  We further employ up to 15 levels of refinement during the course of the simulation which result in an effective spatial resolution of $\sim$ 2000 AU. The same procedure is adopted for all three halos selected from different random seeds. The masses and collapse redshifts of simulated halos are listed in table \ref{tb1}.  A Jeans resolution of at-least 64 cells is ensured during the entire course of simulations, which is found to be sufficient to resolve turbulent eddies in cosmological simulations \citep{Fed11,Latif13c} for which the small scale dynamo gets excited to exponentially amplifiy magnetic fields \citep{Latif13m,LatifMag14}. Our refinement criteria are  based on the baryonic overdensity, the DM mass resolution and  the Jeans refinement, for details see \cite{Latif19,Latif20b}.  We smooth DM particles at level 10 which corresponds to the physical scale of 0.1 pc.  At and below this scale self-gravity from the baryons dominates in the center of the halos and the smoothing of DM particles  avoids a spurious heating of the baryons. 

In the presence of radiative cooling, objects can collapse to very high densities for which the Jeans length cannot be resolved due to the limited resolution and as a result they may undergo artificial fragmentation. To further evolve  the simulations beyond the initial collapse, a minimum pressure is applied to the cells at the maximum refinement level to prevent them from collapsing further and to make them Jeans stable, see \cite{Mach01}.  Artificial pressure is added to the cells at the maximum refinement level taken to be the maximum of $\rm K G \rho_b^2 \Delta_x^2/ \mu$ and the thermal pressure where K is chosen to be 100, $\rm \rho_b$ is the gas density, $\rm \Delta_x$ is the cell width at the maximum level and $\rm \mu$ is the mean particle mass. Doing so may suppress fragmentation over a scale of a few times the cell size at the maximum refinement level which corresponds to a few times 0.01 pc.  Using this technique the Jeans length to cell size ratio is 10. Such an approach has been compared to the sink particles method and was shown to yield similar results \citep{Latif13d, Latif22N}. In our simulations, we employ the non-equilibrium primordial gas chemistry network based on \citet{Abel97} and \citet{Ann97}, to evolve six primordial species (H, H$^+$, He, He$^+$, He$^{2+}$, and e$^-$).  We assume here that the halos are illuminated by an intense Lyman-Werner radiation emitted by nearby sources \citep{Agarwal16,Agarwal19} which suppresses the formation of $\rm H_2$, leading to an isothermal collapse. This enables us to study the impact of magnetic fields on the formation of DCBHs under isothermal conditions.  Our chemical solver is coupled with hydrodynamics and we include cooling from  collisional ionization and excitation  of H and He, radiative recombination, inverse Compton  scattering and bremsstrahlung radiation. 

To identify clumps in our simulations, we employ the clump finder that is publicly available in the YT analysis tool \citep{Turk11a} where there clump finding algorithm is based on \cite{Smith09}. It works by finding topologically connected sets of cells in the spherical region of a given radius. A clump is defined as the mass contained between a local density maximum and the lowest isodensity surface surrounding it.  A clump contains at-least 20  cells and is considered gravitationally bound if its gravitational energy exceeds both the thermal and kinetic energies locally. The algorithm works in density space, the first contour spans the entire range of density within the sphere, effectively creating one large contour which becomes the parent clump of all others to be found later. In the second iteration, contours are created with the same maximum but with the minimum density increased by 1/4 dex. If more than one contours are found then such a group of cells becomes a child clump. This process continues in a recursive manner until the minimum density reaches the maximum one. Effectively, this creates a family tree of clumps and only gravitationally bound child clumps are considered unless no child is found. We take a sphere of radius 10 pc centered at the point of maximum density and set the minimum density to be $\rm 10^{-16} g/cm^3$. We also experimented with further decreasing the minimum density to $\rm 10^{-17} g/cm^3$, increasing the size of the sphere up to 15 pc as well as increasing the minimum number of cells to 32 in the clump and found similar results.

\section{Results} \label{sec:results}

We have performed cosmological magneto-hydrodynamical (MHD) simulations for  three distinct halos of a few times $10^7 ~M_{\odot}$ and compared our results with  hydrodynamical (HD) runs.  For the MHD simulations, we have taken the fiducial value of the initial seed magnetic field to be $\rm 10^{-14}$ G at an initial redshift of 150. This choice of the field is motivated by our previous works L13 and L14. In fact, we found in L14 and  G19 that irrespective of the initial field strength the magnetic field reaches an equipartition value in the presence of strong accretion shocks during the collapse. We also investigated a case with an initial magnetic field strength of $\rm 10^{-20}$ G. In total, we  performed 7 simulations with and without magnetic fields and present their findings here.

\begin{figure*} 
\begin{center}
\includegraphics[scale=1.0]{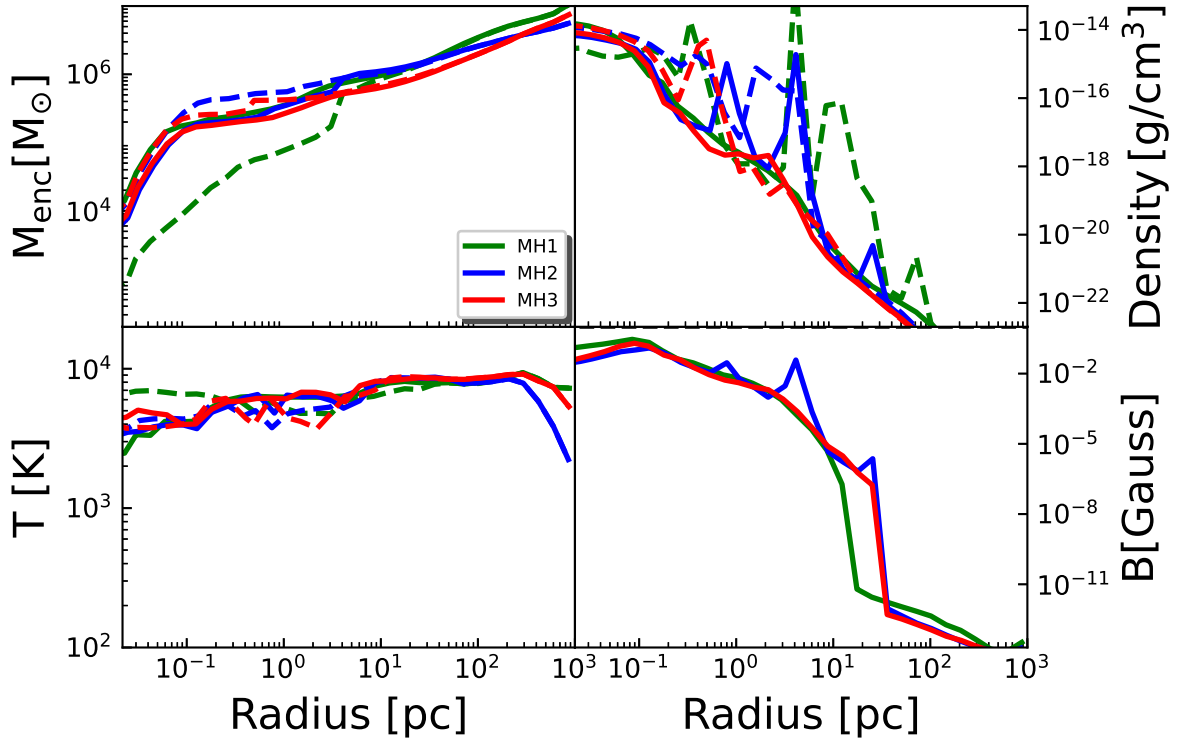}
\end{center}
 \caption{ Average radial profiles of gas density, temperature, enclosed gas mass and magnetic field strength at the end of the simulations. The solid and dashed lines represent the MHD and HD runs, respectively.}
\label{fig1}
\end{figure*}

Gas falls into the dark matter potentials, gets heated due to the virialization shocks and thereafter atomic line cooling kicks in that leads to an isothermal collapse  with T$\sim 8000 $ K in the absence of molecular hydrogen.  Consequently, a monolithic collapse occurs  and the gas densities in the halo center reach $\rm 10^{10}~cm^{-3}$ at scales of about 2000 AU as shown in Fig. \ref{fig1}. The density profiles follow approximately an $\rm R^{-2}$ behavior as expected under  isothermal conditions and the bumps on the profiles reveal the presence of additional clumps at these radii. Moreover, these bumps are more frequent in non-MHD runs  due to the higher degree of fragmentation as we discuss their implications below. Overall, about a few times $\rm 10^6~M_{\odot}$ of gas lies in all simulated halos and small differences in the enclosed gas mass below 10 pc scales are due to fragmentation.
 
\subsection{Magnetic fields amplification}

Magnetic fields get amplified during the gravitational collapse up to a 1 G in the center of the halos as shown in Fig. \ref{fig1}. Initially, the strength of the magnetic field rises slowly but between 20-40 pc it increases by many orders of magnitudes up to $10^{-5}~G$  and then continues to increase towards the halo center.  We observe here that the magnetic field strength increases by 14 orders of magnitude (even 20 for an  initially weak field), while the density increases by 10 orders of magnitude during collapse. This is significantly beyond the amplification expected from compression, where $\rm B \sim \rho ^{2/3}$ under flux freezing.
Any additional contribution is a consequence of rapid amplification by strong accretion shocks and  the small scale dynamo as found in L13, L14 and G19.

\begin{figure*} 
\begin{center}
\begin{tabular}{cc}
\includegraphics[scale=1.0]{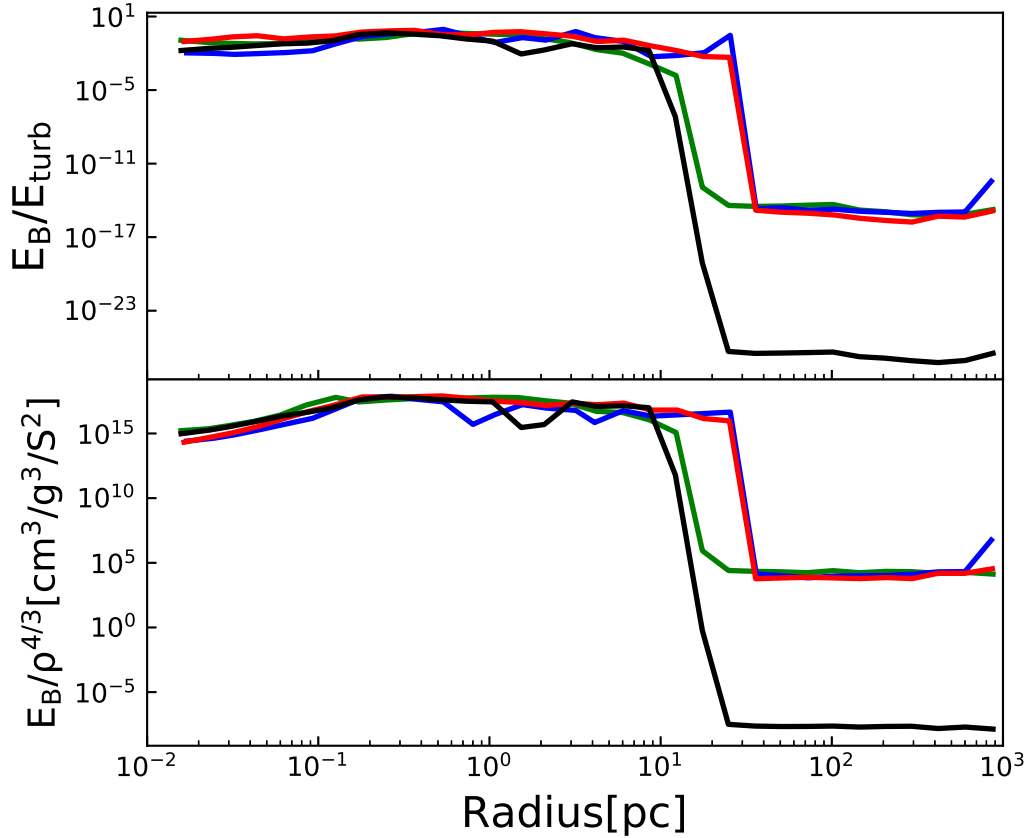}
\end{tabular}
\end{center}
\caption{ The radial profiles show the ratio of magnetic to turbulent energy in the top panel and the magnetic energy normalized by the expected contribution from flux freezing under the assumption of spherically symmetric compression (leading to $\rm E_B \propto \rho^{4/3}$) in the bottom panel. The red, blue and green colors represent the simulations MH1, MH2 \& MH3, while the black solid line represents LH1 with an initial magnetic field strength of $10^{-20}$ G.}
\label{fig2}
\end{figure*}

To  understand the rapid amplification of the magnetic fields, we show the magnetic energy divided by the maximum possible contribution from flux freezing ($\rm \rho^{4/3}$) in Fig. \ref{fig2} assuming a spherical collapse.  The ratio of magnetic energy to $\rm \rho^{4/3}$ shows a big jump between 10-40 pc for all halos and irrespective of the initial magnetic field strength. The magnetic field growth time ($\rm 1/ \omega$) is  about 1000 yrs, which is about an order of magnitude smaller than the collapse timescale. We attribute  this rapid amplification of  magnetic fields  to the strong accretion shocks occurring at these radii which significantly reduce the amplification timescale  and result in a swift amplification of the magnetic fields. Such rapid amplification was investigated in detail by L14 who showed a correlation between the strength of the accretion shocks and rapid amplification, see section 3.1 of L14 and discussion therein.  More recent works by \cite{Hir21} and \cite{Hir22} show similarly swift amplification. Initially the amplification occurs at the edge of the accretion disk and it subsequently grows with time. By the end of the simulations the magnetic field strength spreads outwards  up to 30 pc and becomes  almost constant in the centre.

\begin{figure*} 
\begin{center}
\hspace{-1cm}
\includegraphics[scale=0.76]{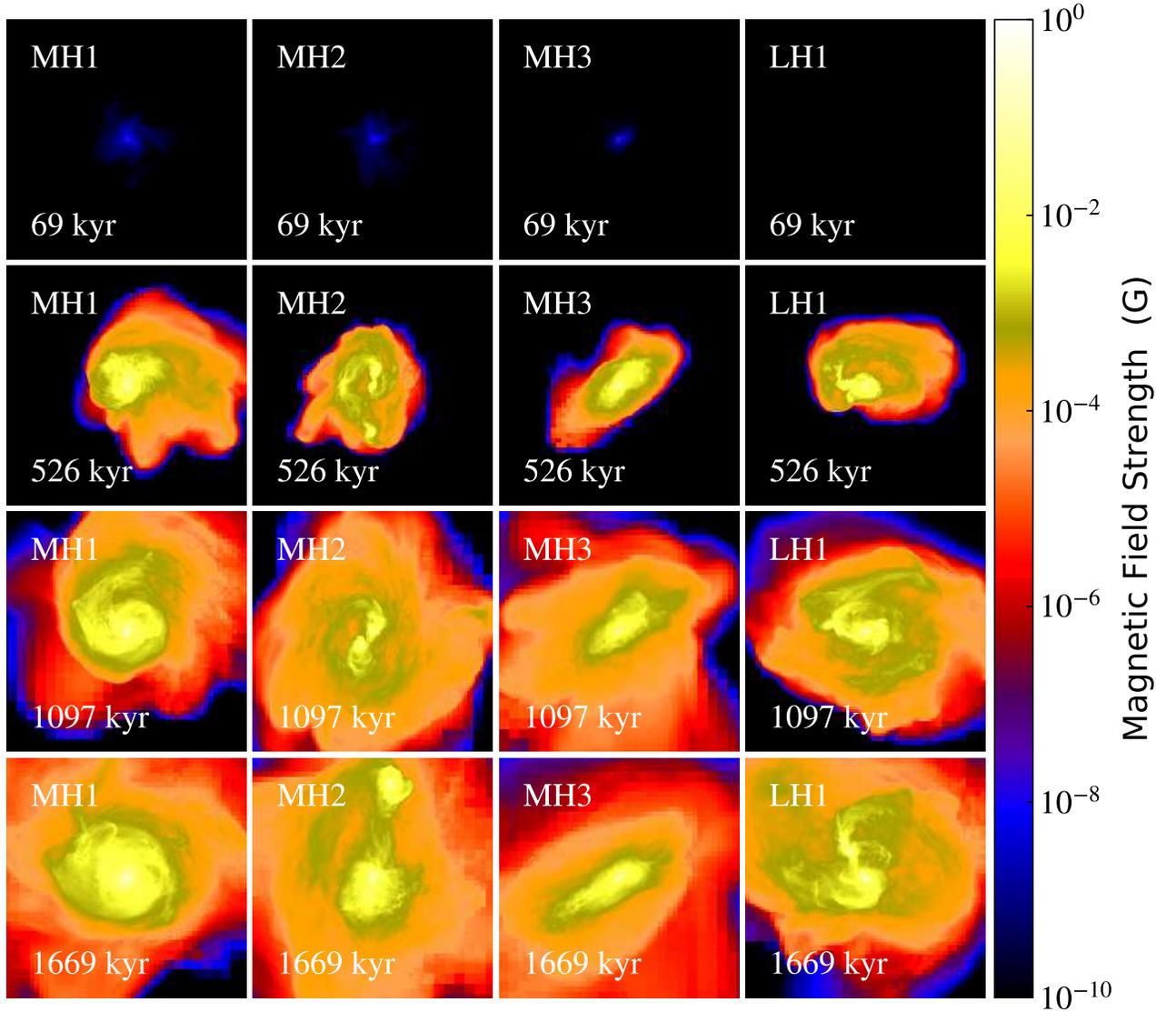}
\end{center}
\caption{The average magnetic field strength along the x-axis is shown here within the central 10 pc. The columns show the time evolution of the magnetic field strengths for the simulations MH1, MH2, MH3 and LH1.}
\label{fig3}
\end{figure*}

To assess the saturation of magnetic fields, we plot the ratio of magnetic to turbulent energy in Fig. \ref{fig2}.  We estimate the turbulent energy as  $\rm E_{turb}=0.5 \rho v_{turb}^2$ and  $\rm v_{turb}=\sqrt{(v_{rms}^2 -v_{rot}^2 - v_{rad}^2)}$ where $\rm v_{rms}$ is the root mean square velocity, $\rm v_{rot} $  is the rotational and $\rm v_{rad}$ is the radial infall velocity. The ratio of turbulent to magnetic energy is almost constant at larger radii,  rapidly increases during strong accretion shocks and reaches close to unity.  This suggests that the magnetic energy is  in equipartition with the turbulent energy and the magnetic field is saturated.  In Fig. \ref{fig3},  we show the time evolution of the magnetic fields strength  in all halos along with the case of the initial magnetic field strength of $\rm 10^{-20}$ G.  It is found that the amplification of the magnetic fields starts in the  centre and their strength increases with time until it gets saturated.  The magnetic field strength is the highest in the center  and expands to  the central 10 pc region in about 1 Myr.  Such strongly amplified and volume filling magnetic fields  favor  the formation of massive black holes as we discuss below.

\subsection{ Impact of magnetic fields on disk fragmentation}

\begin{figure*} 
\begin{center}
\hspace{-1cm}
\includegraphics[scale=0.76]{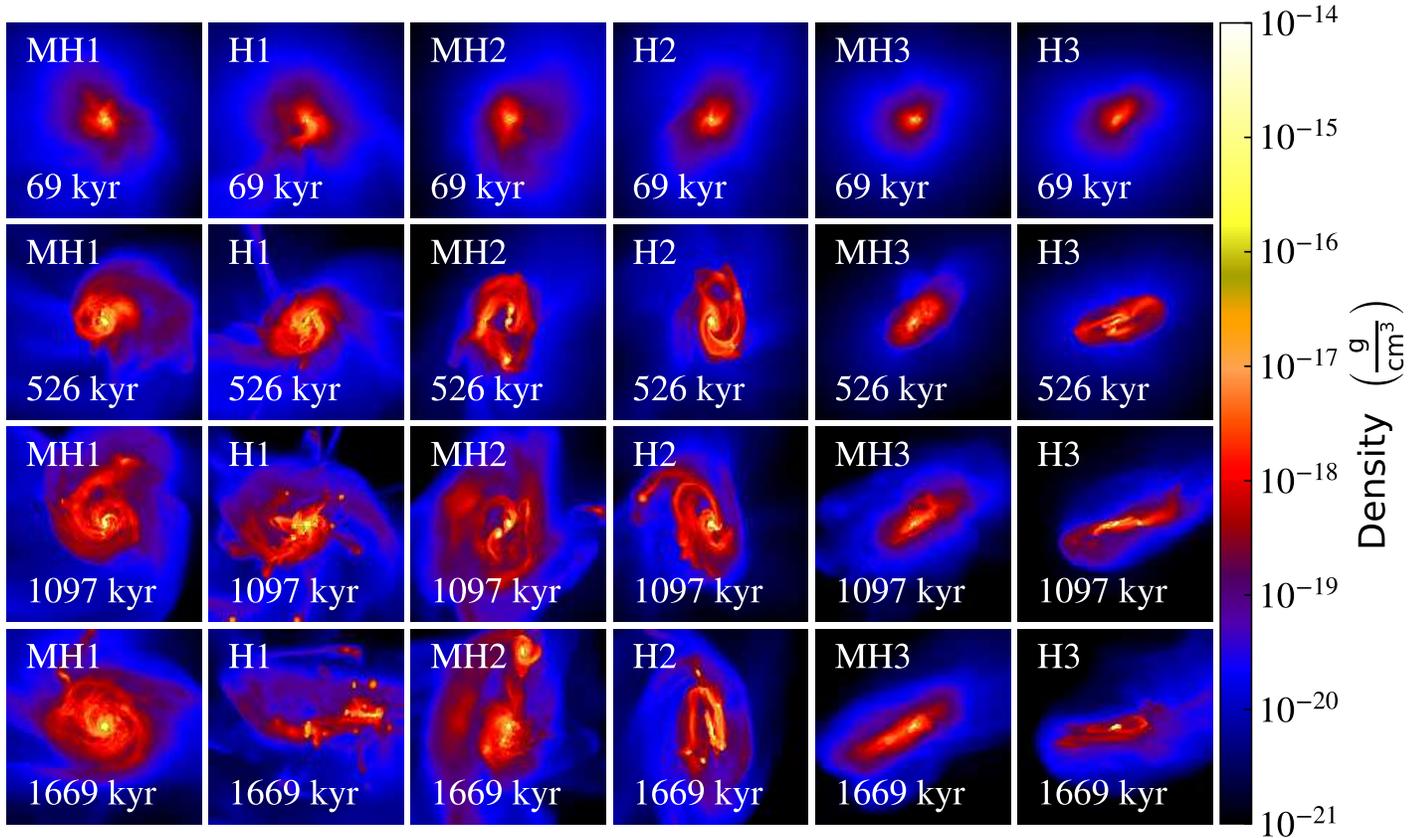}
\end{center}
\caption{The average density along the x-axis is shown here within the central 10 pc for both the MHD and HD runs. The columns show the time evolution of the disk in the MHD and HD simulations.}
\label{fig4}
\end{figure*}

\begin{figure*} 
\begin{center}
\hspace{-1cm}
\includegraphics[scale=0.76]{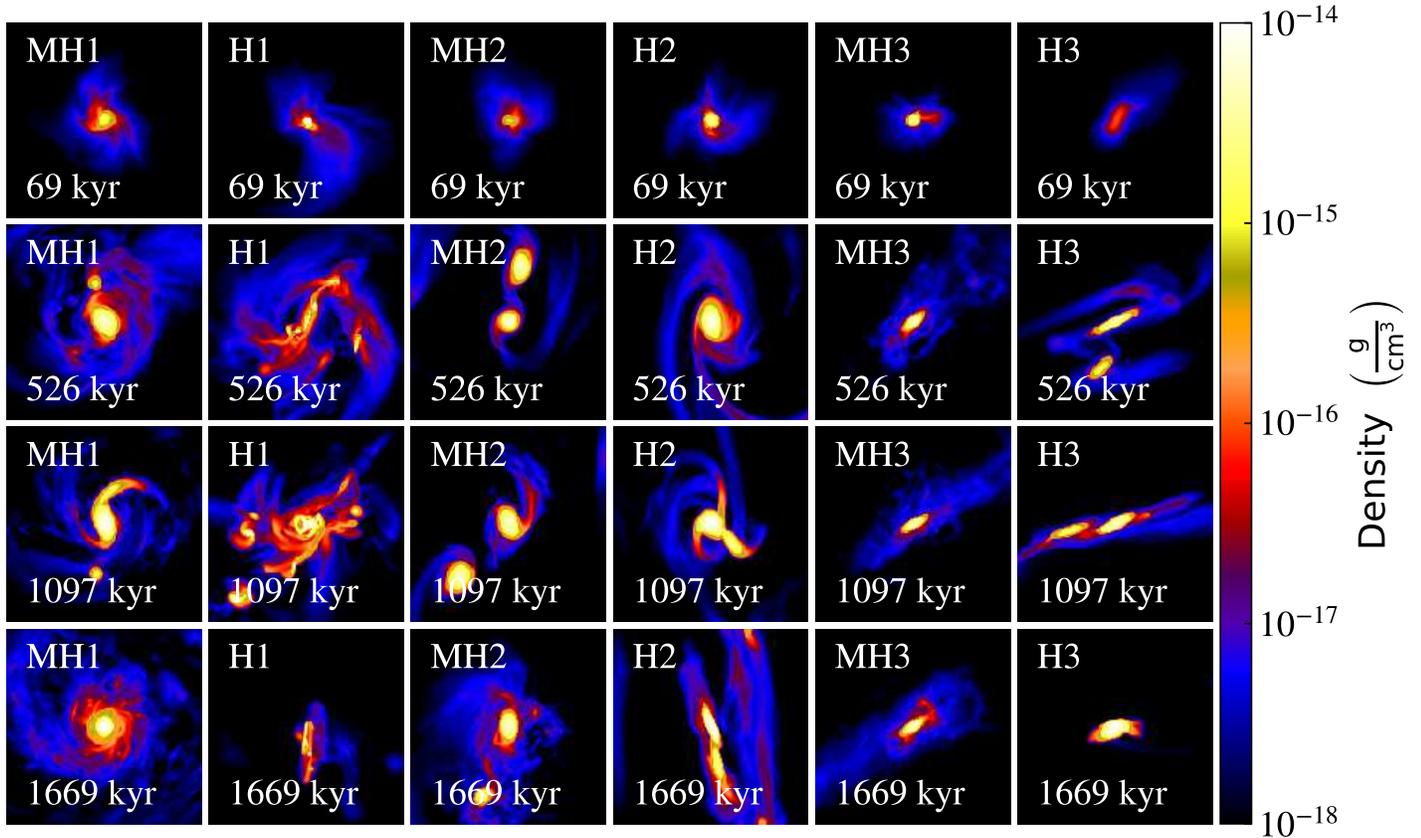}
\end{center}
\caption{Same as Fig. \ref{fig4} except that the central 2 pc view is shown here.}
\label{fig5}
\end{figure*}

We find that  self-gravitating accretion disks form at the center of each halo  as a result of gravitational collapse in both HD and MHD simulations. The time evolution of disks in the simulated halos for both cases  is shown in Fig. \ref{fig4}. In general, the disks are more stable and have larger sizes in the MHD runs while  they are  more compact and gravitationally unstable in HD runs.  Overall, the degree of fragmentation is higher in HD simulations. For MHD runs, no fragmentation  is observed in two out of three halos (halos 1 \&3)  by the end of the simulations as only one clump forming over time gets merged with the central clump. In halo 2,  two clumps are formed and survive for about 1 Myr for the MHD case, see Fig. \ref{fig5} and we expect these clumps to evolve into a binary system.  On the other hand, in the HD runs the disks are more prone to fragmentation and multiple clumps form in the disks, some of them get subsumed in the centre of the disk but a handful of them survive by the end of the simulations.  The differences in fragmentation are due to the magnetic pressure (comparable to the thermal pressure) which increases the Jeans mass and stabilizes the disks.  Consequently,  initially more massive clumps form in MHD runs.

\begin{figure*}
\begin{center}
\begin{tabular}{cc}
\includegraphics[width=0.5\textwidth]{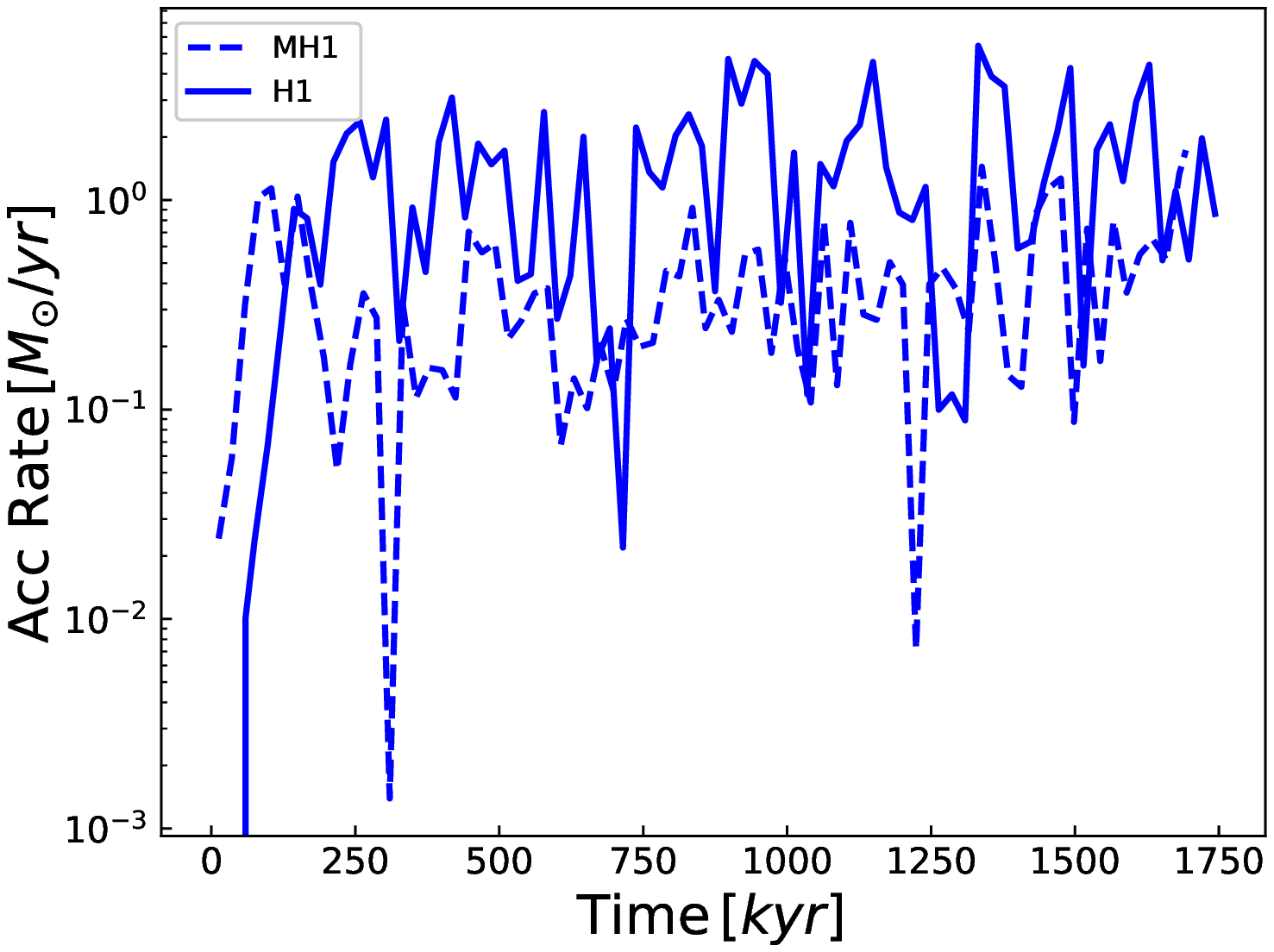} &
\includegraphics[width=0.5\textwidth]{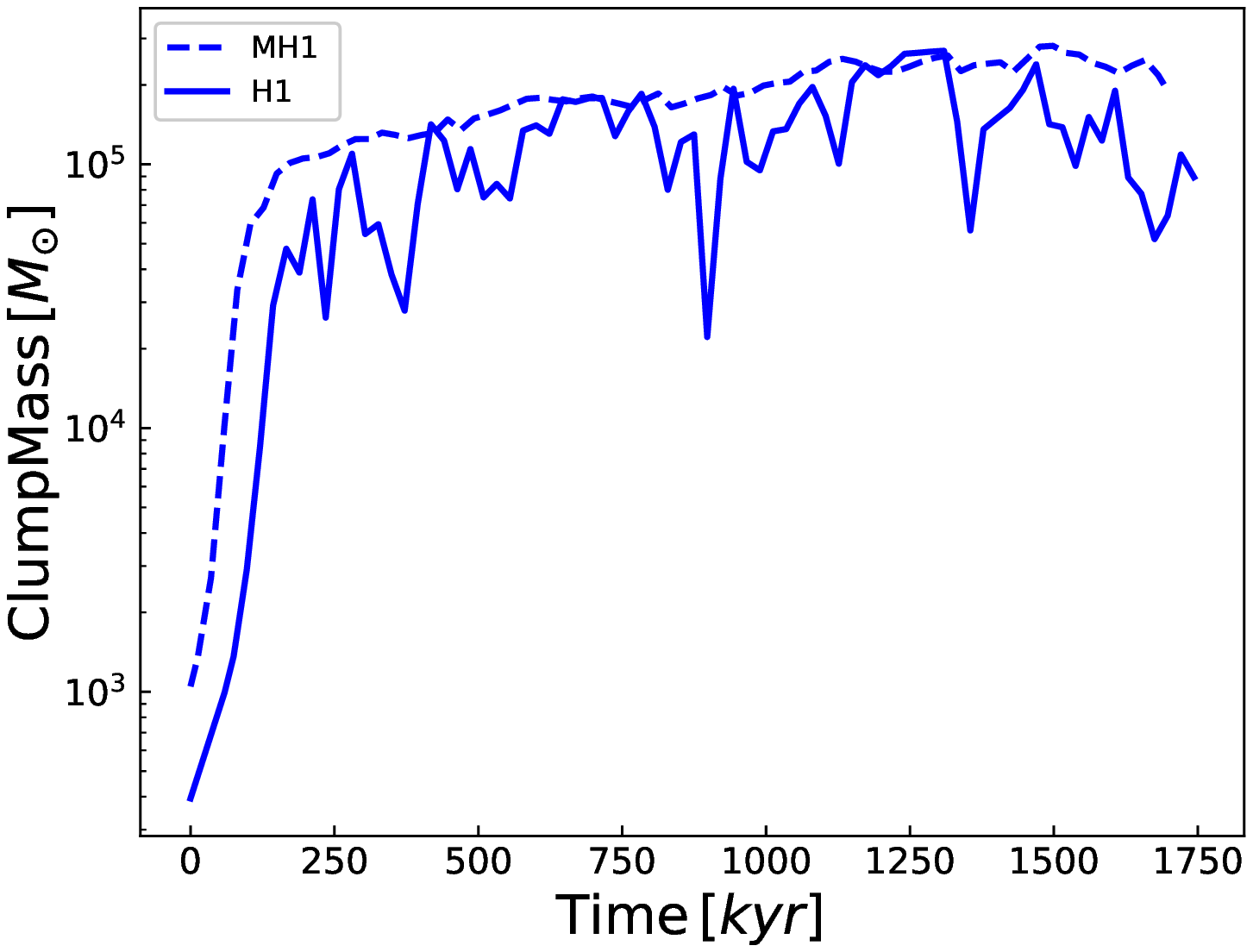} \\
\includegraphics[width=0.5\textwidth]{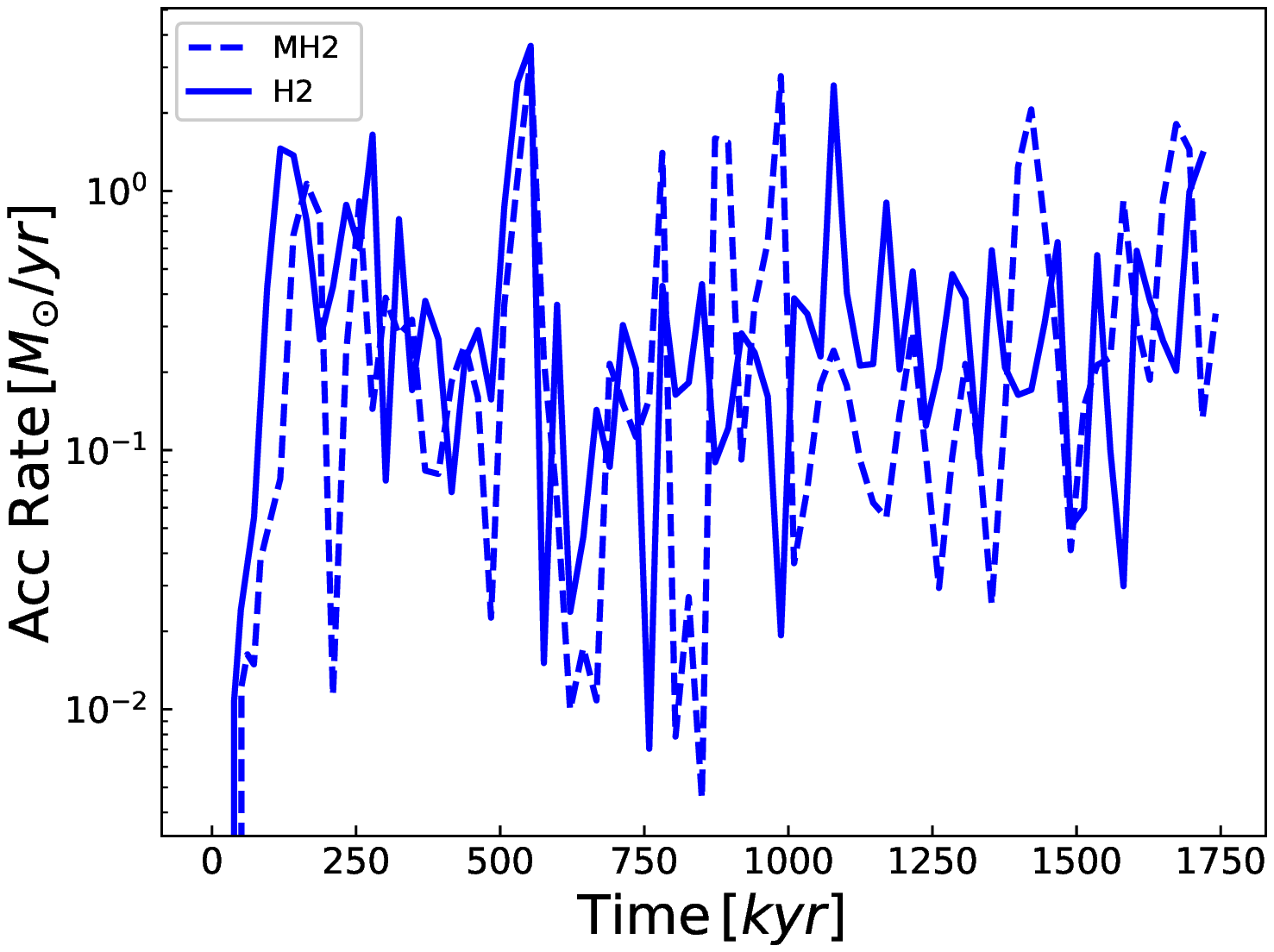} &
\includegraphics[width=0.5\textwidth]{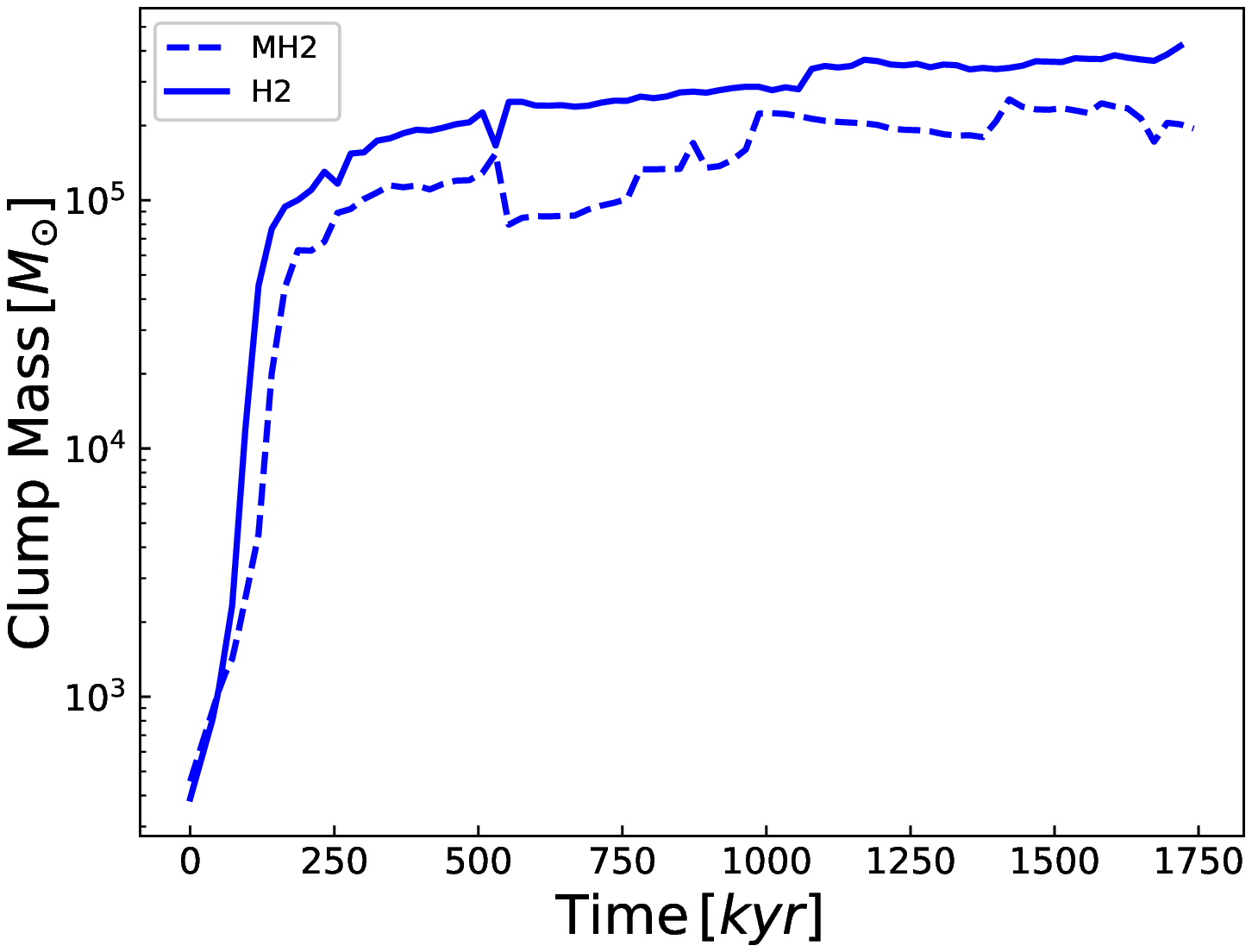} \\
\includegraphics[width=0.5\textwidth]{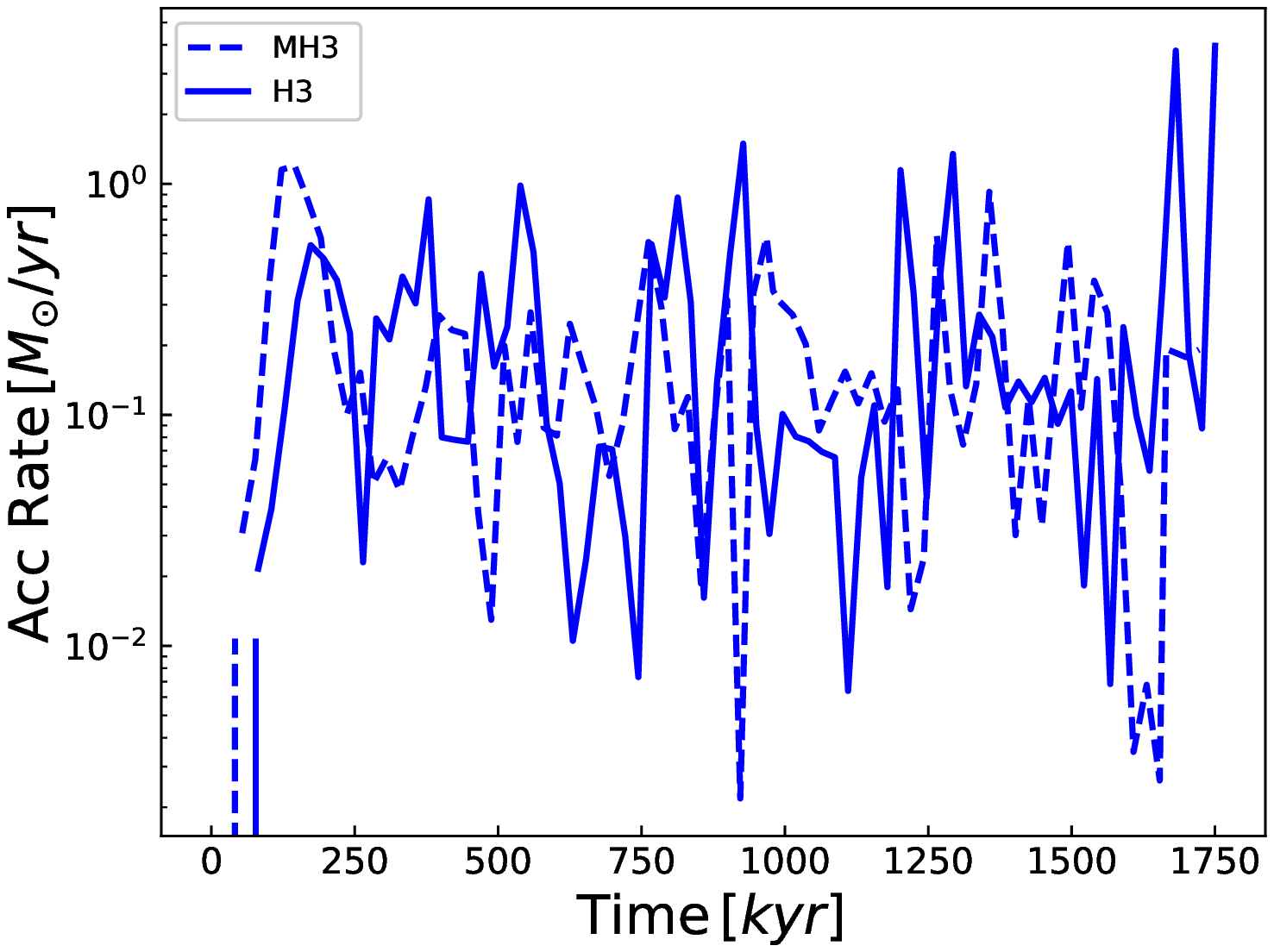} & 
\includegraphics[width=0.5\textwidth]{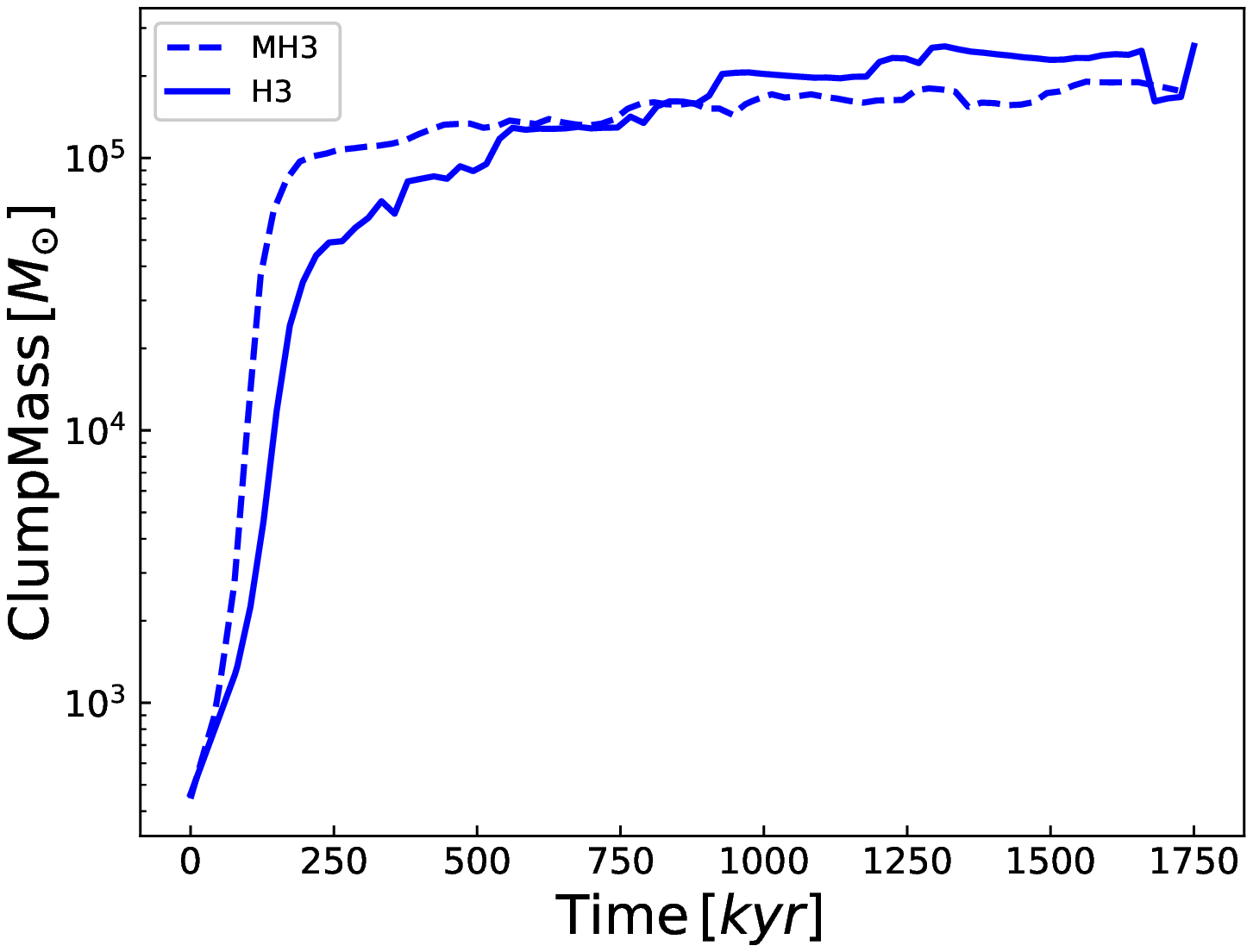} 
\end{tabular}
\end{center}
\caption{ The mass accretion rates for the central 0.1 pc of the halo and the corresponding enclosed masses are shown for the MHD and HD simulations. The solid lines represent the hydrodynamical runs, and the dashed lines the MHD runs.}
\label{fig6}
\end{figure*}

We ran the YT clump finder (see section 2 for details) to estimate the masses of the gravitationally bound clumps. The initial clump masses are estimated when they are found for the first time in data output and are listed in table \ref{tb1} for both MHD and HD runs.  In general, the initial clump masses in HD cases are lower than their MHD counterparts, up to a factor of four in case of halo 2. But by the end of simulations, over a time span of 1.6 Myr, the clump masses in the HD runs become approximately comparable to the ones in MHD as a result of merging.  For MHD cases,  only one clump forms in halos 1 \& 3, two in halo 2 and they have typical masses of   $\rm \sim 10^5~M_{\odot}$. Multiple clumps exist in all halos in the HD runs and their masses range from a few thousands to about a hundred thousand solar. Overall, the central clumps are more massive than the secondary clumps. To comprehend the mass evolution of the central clumps and the differences in the central clump masses between MHD and HD runs, we show the  mass accretion rates  and the masses of the central clumps in Fig. \ref{fig6}. The mass accretion  onto the clumps is  intermittent in both cases and fluctuates about the average accretion rate of $\rm \sim 0.1 ~M_{\odot}/yr$. The peaks in the accretion rates correspond to the epochs  when the clumps merge. In general, the mass accretion rates in MHD runs are  similar to the non-MHD cases. In MH1, the mean accretion rate onto the  main clump is about a factor of 1.5 lower after the initial burst of accretion. 

\begin{figure*} 
\begin{center}
\includegraphics[scale=1.0]{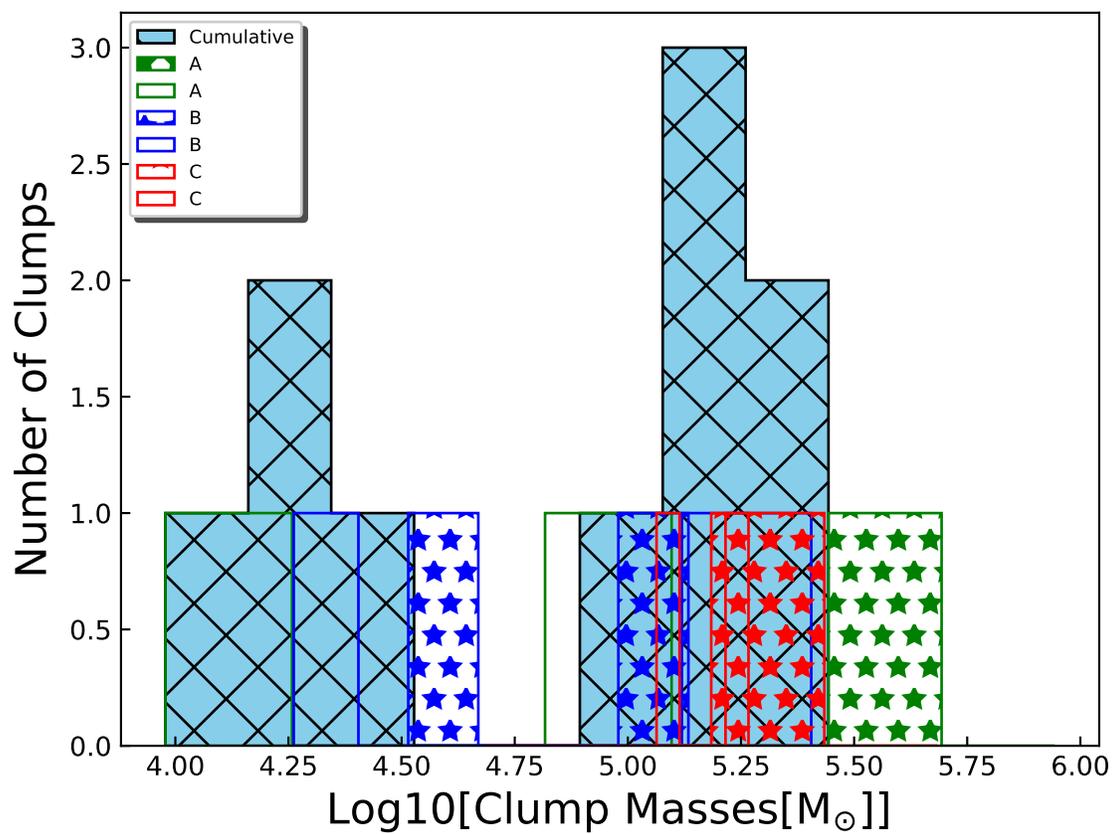}
\end{center}
 \caption{The clump mass distribution at the end of the simulation for both the MHD and HD runs. The filled bars represent the MHD runs and the unfilled bars the HD simulations. The cyan color with crossed hatches shows the cumulative clump mass distribution.}
\label{fig7}
\end{figure*}

\begin{figure*} 
\begin{center}
\includegraphics[scale=1.0]{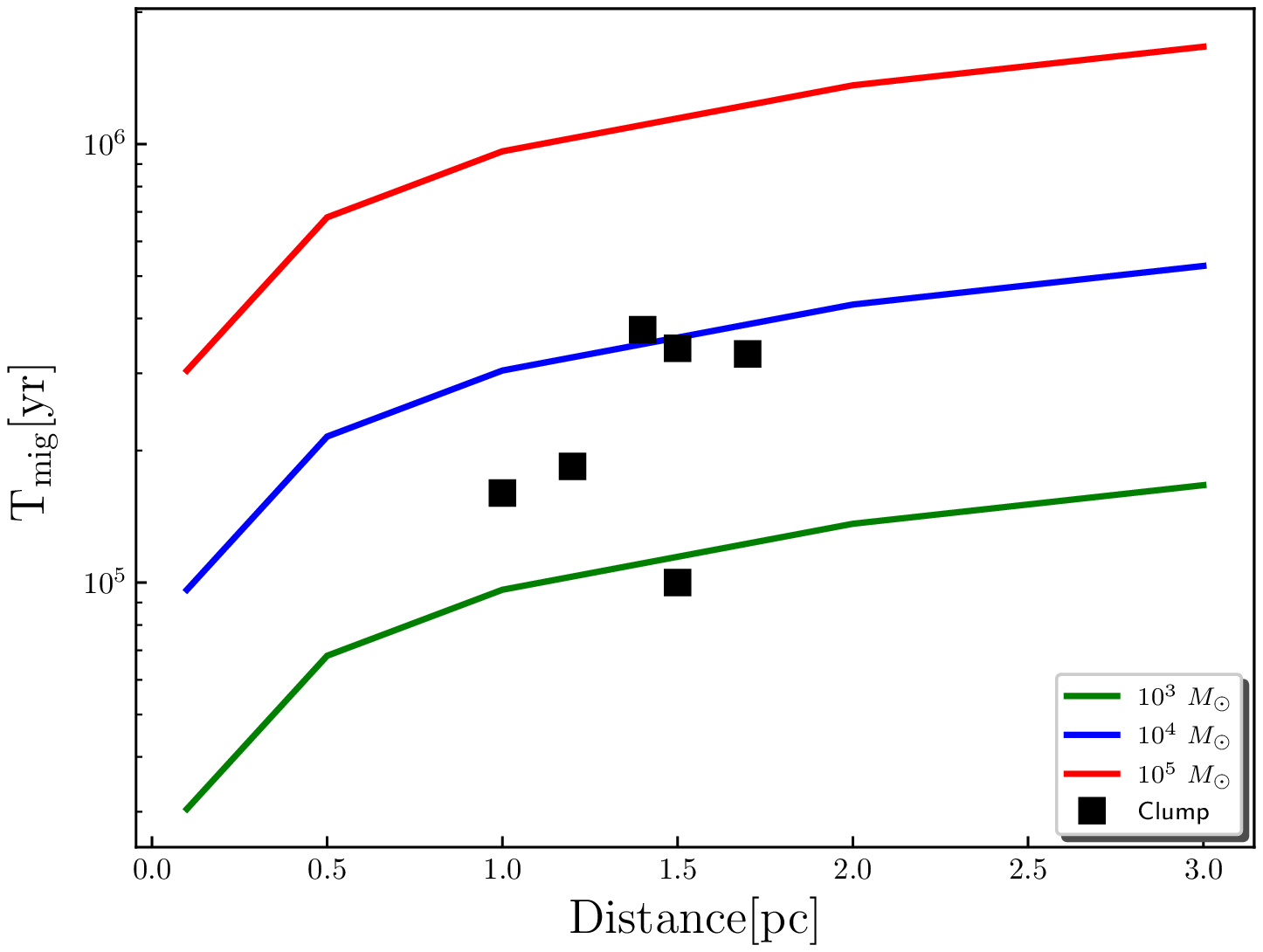}
\end{center}
\caption{ A comparison of the clump migration timescales from numerical simulations with analytic estimates. The solid lines show the analytical results assuming different masses for the central massive object, and the black squares correspond to the simulation data.}
\label{fig8}
\end{figure*}

Although the initial masses of the central clumps  in two out of three halos are  larger in the MHD runs, the merging of secondary clumps in the HD simulations yields similar clump masses at the end of the simulations. For MH1, the initial bursts of accretion boost the  clump mass to 100,000 solar during the first 500 kyrs which gets doubled over  the next Myr.  In spite of  an about 1.5 times higher accretion rate in H1, the formation of multiple clumps and tidal interactions among  them disrupts the growth of central clump and consequently its mass is two times lower than for MH1. The tidal interaction of the clumps is evident from wiggles on the central clump mass plot.  Contrary to MH1, binary clumps form within the first few hundred kyrs in MH2  which result in about factor of  two lower clump masses than the H2. On the other hand in H2,  clumps forming during first 400 kyrs migrate inwards and boost the mass of the central clump which remains two times larger than MHD2 by the end of the simulations. The secondary clumps in H2  form at later times and mainly grow through mergers.  Overall, the mass accretion rate onto the main clumps is similar for both MH2  and H2.  In MH3, although the initial clump mass is a factor of 2-3 larger than H2  during the first 500 kyrs, multiple clumps merge over time in H3 boosting the central clump mass which becomes comparable to the MH3 case.  In general,  the masses of  the main clumps in both MHD and HD cases are $\rm \sim 10^5~M_{\odot}$ and  their mass distribution at the end of the simulations is shown in Fig. \ref{fig7}.  The clump mass distribution is bimodal, the masses of the central clumps in both MHD and HD runs peak at $\rm \sim 10^5~M_{\odot}$ while the masses of the secondary clumps mainly forming in HD runs have masses of $\rm 10^4~M_{\odot}$ except for one of the binary clumps in MH2.
In Fig. \ref{fig8}, we compare the migration timescale of the clumps from simulations with analytical estimates from \cite{Lin86} given in  equation 17 of \cite{Latif5Disk2}. We find that the migration timescales vary from  a few times $10^4 -10^5$ yrs depending on the clump mass and its distance from the centre of a disk.   These estimates are in good agreement with analytical models and explain the migration of  clumps  within 1 Myr after their formation.



\section{Conclusions and Discussion} \label{sec: discussion}

We have investigated  here the impact of magnetic fields on the formation of DCBHs by performing a suite of 3D MHD cosmological simulations for three distinct halos and have compared their results with HD runs.  These simulation are evolved for 1.6 Myrs  comparable to the expected lifetime of SMSs.  For MHD simulations, we have taken a fiducial value of $\rm 10^{-14}$ G for  the initial seed magnetic fields and also explored a case with $\rm 10^{-20}$ G. The simulated halos have masses of $\rm 10^7~M_{\odot}$ at z$\geq$ 12. This study has enabled us to quantify the role magnetic fields during the formation of SMSs.

We found that irrespective of  the initial field strength,  magnetic fields  get rapidly amplified  in the presence of  strong accretion shocks on  times scales shorter than the free-fall time and reach the saturation state.  These results are comparable with previous findings of L14, G19 and \cite{Hir21}. Such strongly amplified magnetic fields enhance the Jeans mass, stabilize the accretion disks and significantly reduce fragmentation. Overall, disks are more stable  and have larger sizes in MHD runs while a higher degree of fragmentation is observed in HD runs.  In spite of  lower initial clump masses in HD simulations, the higher clump merger rate yields  masses of the central clumps similar to the MHD simulations. The masses of  the central clumps  are $\rm \sim 10^5~M_{\odot} $   and the average mass accretion rate   $\rm 0.1 ~M_{\odot}/yr$. They are similar for both the MHD and HD runs. Multiple secondary clumps are found in HD runs by the end of the simulations contrary to the MHD runs.  In halo 2, two clumps have survived for more than 1 Myr and they are expected to form  a binary system.

We have employed here a pressure floor technique which stabilizes the collapse on the smallest scales and allows us to evolve simulations for longer times. This approach suppresses fragmentation on scales of a few times 0.01 pc.  Alternatively sink particles are employed in the  literature which represent gravitationally bound structures. We compared the results of the pressure floor method with sink particles in \cite{Latif13d}  and \cite{Latif22N} and found that both methods yield similar results.  These state of the art methods  are commonly employed in simulations but have their own limitations. The pressure floor technique may increase the minimum size of clumps and consequently enhances the clump merger rate although higher resolution simulations without pressure floors  show that clumps migrate inwards  on shorter times scales and merge with the central clump \citep{Greif11, Latif13c,Sak15,Hos16,Chon18,suaz19}. Also such migration and merging of clumps is in agreement with theoretical models of disk fragmentation \citep{Lin86, Inay14b,Latif5Disk2}. On the other hand contemporary  star formation simulations show that sink creation depends on the choice of the threshold density, numerical resolution, accretion and merging radii as well as on the recipe of sink formation \citep{Gus18,Haugh18}. They further reveal that the sink mass distributions are not well-converged unless simulations have higher spatial resolution, well resolved jeans lengths and smaller merging radii. We expect our main conclusions to remain valid in spite of these caveats.

We have assumed here an isothermal collapse mediated by strong LW radiation. However at densities above $\rm \geq 10^8 cm^{-3}$ radiative association of hydrogen atoms with electrons  results in  photon emission known as H$^-$ cooling. \cite{Latif2016} found that  H$^-$ cooling  decreases the gas temperature down to $\rm \sim$ 5000 K and causes small-scale fragmentation, the clumps at these scales are expected to migrate inwards.  Recently, \cite{Hir21}  included this process in their idealized simulations of atomically cooling gas cloud and  found a higher degree of fragmentation but most of the clumps were able to coalesce with main clump.  Therefore, we expect that the inclusion of H$^-$ cooling will have little impact on our findings but future MHD cosmological simulations should include this process. \cite{ls15} found that magnetic fields can launch Jets and outflows from magnetized SMSs, consistent with duration of gamma ray bursts \citep{Butler18,Sun18} and may even impact the host galaxy.  \cite{wet20a} and  \cite{Wmp21} estimated the radio emission from accreting DCBHs and found that they can be detected with upcoming radio observatories such as SKA and ng-VLA.

\begin{table*}
\begin{center}
\caption{Collapse redshifts, halo masses, initial and final clump masses  are listed here.}
\begin{tabular}{| c | c | c | c| c|c l}
\hline
\hline
Halo & $z$  & Halo Mass  & Initial field strength  & Initial Clump Mass  & Final Clump Mass  \\
    &         &            (\Ms)            &         (Gauss)        &  (\Ms)  &  (\Ms)    \\
\hline
MH1  &  13.2  &  $\rm  3 \times 10^7$  & $\rm 10^{-14} $    &    8751 &  277561 \\
LH1 &    &   & $\rm 10^{-20} $   & 5038 & 169046, 209510  \\
H1 &    &    &     &  5796 & 124926, 88905, 15054, 9508  \\ 
MH2  &  12.15  &  $\rm 1.7 \times 10^7$  & $\rm 10^{-14} $    &  6976 &  136075, 32711  \\
H2  &   &    &     & 1257  & 254245,143027, 18265  \\
MH3  &   12.6 &  $\rm 2.3  \times 10^7$  & $\rm 10^{-14} $   &  4283 & 152833   \\
H3  &    &   &   & 3931 & 184758, 115722 \\

\hline
\end{tabular}

\label{tb1}
\end{center}
\end{table*}

\section{Acknowledgements}

MAL thanks the UAEU for funding via UPAR grant No. 31S390.  DRGS gratefully acknowledges support by the ANID BASAL projects ACE210002 and FB210003, as well as via the Millenium Nucleus NCN19-058 (TITANs). For the purpose of open access, the author has applied a Creative Commons Attribution (CC BY) licence to any Author Accepted Manuscript version arising from this submission. We thank Vanesa Diaz for helpful discussions.
\bibliography{smbhs}
\bibliographystyle{aasjournal}

\end{document}